\documentstyle[epsfig]{lamuphys}
\makeatletter
\let\chapter\hid@chapter
\makeatother

\begin{document}
\pagenumbering{arabic}
\thispagestyle{empty}
\setcounter{page}{0}
\begin{flushright}
UWThPh-1997-39\\
October 1997
\end{flushright}

\vfill

\begin{center}
{\LARGE \bf Pion--Pion and Pion--Nucleon Interactions \\[10pt]
in Chiral Perturbation Theory*}\\[40pt]

\large
Gerhard Ecker\\[0.5cm]
Institut f\"ur Theoretische Physik, Universit\"at Wien\\
Boltzmanngasse 5, A--1090 Wien, Austria \\[10pt]
\end{center}
\vfill

\begin{abstract}
Elastic pion--pion and pion--nucleon scattering are reviewed in the
context of chiral perturbation theory. Theoretical results from
systematic low--energy expansions to $O(p^6)$ for $\pi\pi$ and to
$O(p^3)$ for $\pi N$ are compared with experimental
data. Possible future developments are outlined.
\end{abstract}

\vfill
\begin{center}
Talk given at the Workshop on Chiral Dynamics 1997 \\[5pt] 
Mainz, Germany, Sept. 1 - 5, 1997 \\[5pt]
To appear in the Proceedings
\end{center}

\vfill
\noindent * Work supported in part by FWF (Austria), Project
No. P09505--PHY and by HCM, EEC--Contract No. CHRX--CT920026
(EURODA$\Phi$NE). 
\newpage

 \title{Pion--Pion and Pion--Nucleon
Interactions\\ in Chiral Perturbation Theory}

\author{Gerhard\,Ecker}

\institute{Inst. Theor. Physik, Universit\"at Wien, Boltzmanng. 5,
A-1090 Wien, Austria}

\maketitle

\begin{abstract}
Elastic pion--pion and pion--nucleon scattering are reviewed in the
context of chiral perturbation theory. Theoretical results from
systematic low--energy expansions to $O(p^6)$ for $\pi\pi$ and to
$O(p^3)$ for $\pi N$ are compared with experimental
data. Possible future developments are outlined.
\end{abstract}
\section{Overview}
Elastic $\pi\pi$ and $\pi N$ scattering are the classical scattering
processes of hadron physics, with a long history both in experiment
and in theory.  The purpose of this talk is to review the present
status of these processes within the effective field theory of the
standard model at low energies, chiral perturbation theory (CHPT)
(Weinberg 1979; Gasser and Leutwyler 1984, 1985; Leutwyler 1994).

With pions and nucleons only, chiral $SU(2)$ is the appropriate
setting.  For this review, isospin conservation is assumed and the
electromagnetic interaction is turned off. The framework will be
standard CHPT (Gasser and Leutwyler 1984, 1985), but I will contrast
the standard CHPT calculations for $\pi\pi$ scattering (Bijnens et
al. 1996, 1997) with the extensive work performed in the framework of
generalized CHPT (Knecht et al. 1995, 1996). For elastic $\pi N$
scattering, the only complete calculation to $O(p^3)$ (Moj\v zi\v s
1997) is based on the standard scheme.

Comparing the two reactions from the chiral point of view, the
differences are more apparent than the similarities. Like any
non--Goldstone degrees of freedom, the nucleons are less restricted by
chiral symmetry than the pions. In addition, the chiral expansion
proceeds in steps of $p$ in the pion--nucleon case rather than $p^2$
in the purely mesonic case, with $p$ as usual a generic small
momentum. Both facts explain why the presently available amplitudes
involve about the same number of low--energy constants for the two
processes, even though we have reached $O(p^6)$ for $\pi\pi$, but only
$O(p^3)$ for $\pi N$.  The $\pi\pi$ amplitude is now known up to and
including the two--loop level. The $\pi N$ amplitude, on the other
hand, is still not completely known even to one--loop accuracy as long
as the $O(p^4)$ part is missing.

For chiral $SU(2)$, one may expect good convergence (in the usual
sense of CHPT) of the low--energy expansion near threshold.
Anticipating the summary, the convergence is by now very satisfactory
for $\pi\pi$, but not sufficient yet for $\pi N$. This is in contrast
to the experimental situation. The available data for $\pi\pi$
scattering near threshold are scarce and not precise enough for
definitive tests of the CHPT
predictions. However, much of the recent theoretical activity in this
field has been motivated by the experimental prospects for significant
improvements in the near future, at DA$\Phi$NE (Lee-Franzini 1997),
BNL (Lowe 1997) and CERN (Schacher 1997). For elastic $\pi N$
scattering, on the other hand, data are abundant. Here the precision
seems almost too good for some of the quantities extracted from
experiment, given the inconsistencies in the existing data
sample (H\"ohler 1997; Pavan 1997).

\section{Pion--Pion Scattering}
In the isospin limit, the scattering amplitude for
\begin{equation}
\pi^a(p_1) + \pi^b(p_2) \to \pi^c(p_3) + \pi^d(p_4)
\end{equation}
is determined by a single scalar function $A(s,t,u)$ of the usual
Mandelstam variables $s,t,u$ :
\begin{eqnarray}
T_{ab,cd}(s,t,u) &=& \delta_{ab}\delta_{cd} A(s,t,u) +
\delta_{ac}\delta_{bd} A(t,s,u) + \delta_{ad}\delta_{bc} A(u,t,s)
\nonumber \\ A(s,t,u) &=& A(s,u,t) ~.
\end{eqnarray}

The amplitudes of definite isospin $(I = 0,1,2)$ in the $s$--channel
are decomposed into partial--wave amplitudes $t_l^I(s)$. In the
elastic region $4M_\pi^2 \leq s \leq 16M_\pi^2$, the partial--wave
amplitudes can be described by real phase shifts $\delta_l^I(s)$:
\begin{equation}
t_l^I(s)=(1-\displaystyle \frac{4 M_\pi^2}{s})^{-1/2} \exp{i
\delta_l^I(s)} \sin{\delta_l^I(s)}~.
\end{equation}
The behaviour of the partial waves near threshold is of the form
\begin{equation}
\Re e\;t_l^I(s)=q^{2l}\{a_l^I +q^2 b_l^I +O(q^4)\}~,
\label{effrange}
\end{equation}
with $q$ the center--of--mass momentum.  The quantities $a_l^I$ and
$b_l^I$ are referred to as scattering lengths and slope parameters,
respectively.

The amplitude $A(s,t,u)$ was calculated to $O(p^2)$ by \cite{wein66}
and to $O(p^4)$ by \cite{glplb}. To next--to--next--to--leading order,
$O(p^6)$, the amplitude is now available in two different forms as
described in the following two subsections.

\subsection{Dispersive Calculation}
Unitarity for the partial--wave amplitudes,
\begin{equation}
\Im m\;t_l^I(s)=(1-\displaystyle \frac{4 M_\pi^2}{s})^{1/2}
|t_l^I(s)|^2 + \mbox{inelastic contributions}\; (\mbox{for } s > 16
M_\pi^2)~,
\label{unitary}
\end{equation}
leads to the following consequences:
\begin{description}
\item[i.]  If $t_l^I(s)$ is known to $O(p^{2n})$, $\Im m\;t_l^I(s)$
can be calculated in the elastic region to $O(p^{2n+2})$ from the
unitarity relation (\ref{unitary}).
\item[ii.]  Since the lowest--order amplitude of $O(p^2)$ corresponds
to partial waves with $l=0,1$ only, Eq.~(\ref{unitary}) implies
\begin{equation}
\Im m\;t_l^I(s)=O(p^8) \qquad \mbox{for} \; l\ge 2~.
\end{equation}
\end{description}
Inelastic contributions enter at $O(p^8)$ only.

Given the amplitude to $O(p^4)$, unitarity and analyticity therefore
allow for a dispersive calculation (Knecht et al. 1995) of $A(s,t,u)$
to $O(p^6)$ up to a crossing symmetric subtraction polynomial. The
analytically nontrivial part has a relatively simple form (Stern et
al. 1993) that can be expressed in terms of up to third powers of the
standard one--loop function. The subtraction polynomial depends on six
parameters four of which can be obtained from sum rules involving
high--energy $\pi\pi$ data (Knecht et al. 1996).

The general form of the scattering amplitude given by \cite{KMSF95} is
valid for both the standard and the generalized (Stern 1997)
picture. The differences are all contained in the six subtraction
constants.

\subsection{Field Theoretic Calculation}
The diagrammatic calculation of $A(s,t,u)$ to $O(p^6)$ (Bijnens et
al. 1996, 1997) in the standard framework is quite involved. The main
features for a comparison with the dispersive approach are the
following:
\begin{itemize}
\item
The analytically nontrivial part of the dispersive result is
reproduced as well as the general structure of the polynomial piece.
\item
To arrive at the final renormalized amplitude, one needs in addition
the following quantities to $O(p^6)$: the pion wave function
renormalization constant (B\"urgi 1996), the pion mass (B\"urgi 1996)
and the pion decay constant (Bijnens et al. 1996, 1997).
\item
In the notation of Bijnens et al. (1996, 1997), the subtraction
polynomial is expressed in terms of six dimensionless parameters $b_i$
($i=1,\dots,6$).  The field theoretic approach produces these
parameters as functions
\begin{equation}
b_i(M_\pi/F_\pi,M_\pi/\mu;l_i^r(\mu),k_i^r(\mu))~,
\end{equation}
where $\mu$ is an arbitrary renormalization scale, the $l_i^r$
($i=1,\dots,4$) are the relevant low--energy constants of $O(p^4)$
(Gasser and Leutwyler 1984) and the $k_i^r$ are six combinations of
the corresponding constants of $O(p^6)$ in the $SU(2)$ version of the
chiral Lagrangian of \cite{fs96}.
\end{itemize}

Compared to the dispersive approach, the diagrammatic method offers
the following advantages:
\begin{description}
\item[i.]  The full infrared structure is exhibited to $O(p^6)$. In
particular, the $b_i$ contain chiral logs of the form
$(\ln{M_\pi/\mu})^n$ ($n\le 2$) that are known to be numerically
important, especially for the infrared--dominated parameters $b_1$ and
$b_2$. On the other hand, $b_3$, \dots, $b_6$ are more sensitive to
the ``high--energy'' structure. The latter are precisely the four
parameters for which sum rule estimates exist (Knecht et al. 1996).
\item[ii.]  The explicit dependence on low--energy constants makes
phenomenological determinations of these constants and comparison with
other processes possible. This is especially relevant for determining
$l_1^r$, $l_2^r$ to $O(p^6)$ accuracy (Colangelo et al. 1997b).
\item[iii.]  The fully known dependence on the pion mass allows one to
evaluate the amplitude even at unphysical values of the quark mass
(remember that we assume $m_u=m_d$). One possible application is
to confront the CHPT amplitude with (unquenched) lattice calculations
(Colangelo 1997a).
\end{description}

\subsection{Results}
In the standard picture, the $\pi\pi$ amplitude depends on four
low--energy constants of $O(p^4)$ and on six combinations of $O(p^6)$
couplings. The latter have been estimated with meson resonance
exchange that is known to account for the dominant features of the
$O(p^4)$ constants (Ecker et al. 1989).  Referring to \cite{BCEGS2}
for details, the inherent uncertainties of this approximation induce
small (bigger) uncertainties for the low (higher) partial waves. The
main reason is that the higher partial waves are more sensitive to the
high--energy structure.

Concerning the low--energy constants of $O(p^4)$, the $\pi\pi$
amplitude of $O(p^6)$ will lead eventually to a more precise
determination of some of those constants (Colangelo et al. 1997b) than
presently available. For the time being, one can investigate the
sensitivity of the amplitude to the $\bar{l}_i$, the scale independent
couplings introduced by \cite{glann}. One obvious choice is based on
the original analysis to $O(p^4)$ (Gasser and Leutwyler 1984)
supplemented by a more recent investigation of $K_{e4}$ decays
including dispersive estimates of higher--order effects (Bijnens et
al. 1994). The following values are referred to as set I :
\begin{eqnarray}
\begin{array}{rrrr}
\bar{l}{_1}=&-1.7 \; \; ,& \bar{l}{_2}=& 6.1 \; \; , \\ \bar{l}{_3}=&
2.9\; \; ,& \bar{l}{_4}=& 4.3 \;\; .
\end{array} \label{setI}
\end{eqnarray}
As first emphasized by \cite{girl}, the amplitude for set I leads to
$D$--wave scattering lengths that do not agree well with the values
extracted from experiment (see Table \ref{tab:thresh} below). For set
II we have therefore updated the procedure of \cite{glann} to fix
$\bar{l}_1$, $\bar{l}_2$ from the $D$--wave scattering lengths
$a_2^0$, $a_2^2$, now to $O(p^6)$ accuracy:
\begin{eqnarray}
\begin{array}{rrrr}
\bar{l}{_1}=&-1.5 \; \; ,& \bar{l}{_2}=& 4.5 \;\; ,
\end{array} \label{setII}
\end{eqnarray}
leaving $\bar{l}_3$, $\bar{l}_4$ unchanged. Although $\bar{l}_1$ is
practically unchanged, the associated error is large because the
$D$--wave scattering lengths depend only weakly on $\bar{l}_1$. On the
other hand, the decrease of $\bar{l}_2$ from set I to set II is more
pronounced. In fact, there are some independent indications in favour
of such a smaller value of $\bar{l}_2$ (Pennington and Portol\'es
1995; Ananthanarayan and B\"uttiker 1996; Wanders 1997).

The dependence on the low--energy constants is contained in the
parameters $b_i$. It turns out (Bijnens et al. 1997) that $b_1$ and
$b_2$ are rather insensitive to the precise values of the $\bar{l}_i$
whereas $b_3$,\dots, $b_6$ change substantially between sets I and
II. This is of course in line with the previous observation that
$b_1$, $b_2$ are infrared dominated while the other $b_i$ are more
sensitive to the high--energy structure.

\begin{figure}[t]
\begin{center}
\mbox{\epsfysize=8cm \epsfbox{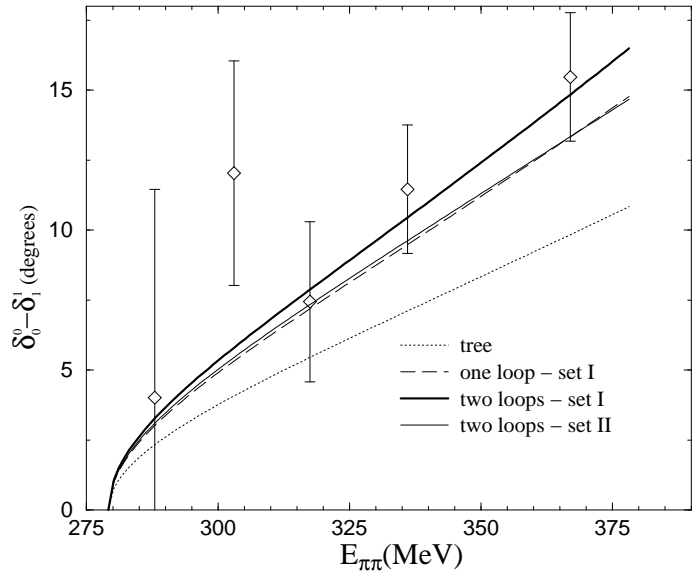} }
\caption{Phase shift difference $\delta_0^0-\delta_1^1$ at $O(p^2)$,
  $O(p^4)$ and $O(p^6)$ (set I and II) from Bijnens et al. (1997).}
\label{fig:d00md11}
\end{center}
\end{figure}

\begin{figure}[htb]
\begin{center}
\mbox{\epsfysize=8cm \epsfbox{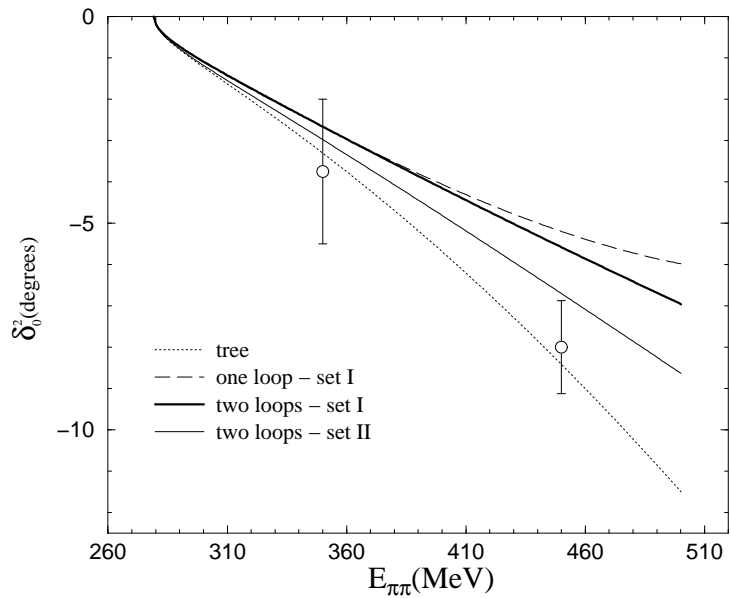} }
\caption{Phase shift $\delta_0^2$ at $O(p^2)$, $O(p^4)$ and $O(p^6)$
  (set I and II) from Bijnens et al. (1997). }
\label{fig:d20}
\end{center}
\end{figure}

In Figs. \ref{fig:d00md11}, \ref{fig:d20} the phase shift difference
$\delta_0^0 - \delta_1^1$ and the $I=2$ $S$--wave phase shift
$\delta_0^2$ are plotted as functions of the center--of--mass energy
and compared with the available low--energy data.  The two--loop phase
shifts describe the $K_{e4}$ data (Rosselet et al. 1977) very well for
both sets of $\bar{l}{_i}$, with a small preference for set I. The
$I=2$ $S$ wave, on the other hand, seems to prefer set II.


The lower value of $\bar{l}_2$ in set II raises the question whether
such a value is still compatible with the idea of resonance saturation
of the low--energy constants of $O(p^4)$. To illustrate this point, I
take the resonance dominance values for the $l_i^r(\mu)$ as given in
\cite{EGPR89} and choose the renormalization scale $\mu=M_\eta$. This
gives rise to set III:
\begin{eqnarray}
\begin{array}{rrrr}
\bar{l}{_1}=&-0.7 \; \; ,& \bar{l}{_2}=& 5.0 \; \; , \\ \bar{l}{_3}=&
1.9\; \; ,& \bar{l}{_4}=& 3.7 \;\; .
\end{array} \label{setIII}
\end{eqnarray}

\begin{table}[ht]
\caption{Threshold parameters in units of $M_{\pi^+}$ for three sets
of low--energy constants $\bar{l}{_i}$ defined in
Eqs.~(\protect\ref{setI},\protect\ref{setII},\protect\ref{setIII}).
The values of $O(p^4)$ correspond to set I.  The experimental values
are from \protect\cite{threshexp}.}
\label{tab:thresh}

\begin{center}
\begin{tabular}{c||c|c|c|c|c|c}
\hline & & & & & & \\ &\mbox{ } $O(p^2)$\mbox{ } & \mbox{ } $O(p^4)$
\mbox{ }& \mbox{ } $O(p^6)$ \mbox{ }&\mbox{ } $O(p^6)$ \mbox{ }&
\mbox{ } $O(p^6)$ \mbox{ } &\mbox{ } experiment \mbox{ } \\ & & & set
I & set II & set III & \\[2pt] \hline $ a_0^0$& $0.16$ & $0.20$ &
$0.217$ & $0.206$ & $0.209$ & $0.26\pm0.05$\\

$b_0^0$ & $0.18$ & $0.25$& $0.275$ &$0.249$ & $0.261$ &
$0.25\pm0.03$\\

$-$10 $a_0^2$ & $0.45$ & $0.42$ &$0.413$ & $0.443$ & $0.415$ &
$0.28\pm0.12$\\

$-$10 $b_0^2$ & $0.91$ & $0.73$ &$0.72$ & $0.80$ & $0.75$ &
$0.82\pm0.08$\\

10 $a_1^1$ & $0.30$ & $0.37$ & $0.40$ & $0.38$ & $0.37$ &
$0.38\pm0.02$\\

$10^2 b_1^1$ & $0$ & $0.48$ & $0.79$ & $0.54$ & $0.60$ & \\

$10^2 a_2^0$ & $0$ & $0.18$ & $0.27$ & input & $0.19$ & $0.17\pm0.03$
\\

$10^3 a_2^2$ & $0$ & $0.21$ & $0.23$ & input & $0.16$ & $0.13\pm0.30$
\\ \hline
\end{tabular}
\end{center}
\end{table}

In Table \ref{tab:thresh}, the threshold parameters for the low
partial waves are displayed at $O(p^2)$, $O(p^4)$ (set I) and to
$O(p^6)$ for the three sets I, II and III. The experimental values are
taken from the compilation of \cite{threshexp}.  Referring again to
\cite{BCEGS2} for additional discussion, let me emphasize two relevant
points here:
\begin{itemize}
\item
The $S$--wave threshold parameters are very stable, especially
the $I=0$ scattering length, while the higher partial waves are 
more sensitive to the choice of low--energy constants of $O(p^4)$ 
(and of $O(p^6)$, for that matter).
\item
The resonance dominance prediction (set III) is in perfect agreement
with the data although the agreement becomes less impressive for $\mu
> M_\eta$.
\end{itemize}

Table \ref{tab:thresh} also documents that more work is needed to
extract the low--energy constants from the data and to assign credible
theoretical errors to phase shifts and threshold parameters.  Such an
analysis based on the Roy equation machinery (Roy 1971) is well under
way (Colangelo et al. 1997b). A similar approach has been used by
\cite{KMSF96} to determine the parameters $b_3$,\dots, $b_6$ via sum
rules.

Without anticipating the results of such an analysis, it is clear from
Table \ref{tab:thresh} that the $I=0$ $S$--wave scattering length is
very stable with respect to reasonable changes of the $\bar{l}_i$ and
that a value
\begin{equation}
a_0^0 = 0.21 \div 0.22
\label{eq:a00}
\end{equation}
is strongly favoured. Moreover, as shown in Fig.~\ref{fig:d00md11},
the corresponding phase shift gives an excellent description of the
$K_{e4}$ data, especially for set I (the curve for set III not shown
explicitly lies between those for sets I and II). This makes one
wonder why the mean value of the usually quoted $a_0^0=0.26\pm 0.05$
(Froggatt and Petersen 1977) is substantially bigger than
(\ref{eq:a00}), although of course consistent within the error.

To make the point that with the present experimental precision $a_0^0$
depends crucially on the fitting procedure, I analyze the data of
\cite{ross} in Fig.~\ref{fig:d00md11} in two different ways. The first
method is the one employed by the experimentalists themselves who used
the approximate formula
\begin{equation}
\sin{2(\delta_0^0-\delta_1^1)} = 2 \sqrt{1-\displaystyle \frac{4
M_\pi^2}{s}} \; (a_0^0 + q^2 b)~.\label{eqross}
\end{equation}
Fitting the data with free parameters $a_0^0$ and $b$ produces a mean
value $a_0^0 = 0.31$.  Employing a relation between $a_0^0$ and $b$
attributed to \cite{BFP74}, the final value of \cite{ross} is
$a_0^0=0.28\pm0.05$. The second method for extracting $a_0^0$ uses the
full CHPT amplitude to $O(p^6)$. To perform a one--parameter fit, I
choose for definiteness set I but leave $\bar{l}_2$ as a free
parameter. The fit has practically the same $\chi^2$ as the previous
one and gives rise to
\begin{eqnarray}
a_0^0 &=& 0.220 \pm 0.012 \nonumber \\ \bar{l}{_2}&=& 6.4 \pm 1.6~.
\label{fitB}
\end{eqnarray}

The main reason for the different values of $a_0^0$ extracted from the
same data is that the scattering length only dominates in a small
neighbourhood of the threshold (Leutwyler 1997). Even for the limited
range in $E_{\pi\pi}$ covered by the experiment of \cite{ross},
higher--order terms in the threshold expansion (\ref{effrange}) are
important. It goes without saying that the full chiral amplitude to
$O(p^6)$ has a superior theoretical status than the simple
approximation (\ref{eqross}) on all accounts. Therefore, I conclude
that the CHPT prediction (\ref{eq:a00}) is in perfect agreement with
the available data and that there is no indication for an unusually
small quark condensate (Stern 1997) on the basis of existing $\pi\pi$
data.  By no means, this is meant to imply that we do not need more
precise experimental information to really pin down the quark
condensate from $\pi\pi$ scattering (Lee-Franzini 1997; Lowe 1997;
Schacher 1997).

\section{Pion--Nucleon Scattering}
The amplitude for pion--nucleon scattering
\begin{equation}
\pi^a(q_1) + N(p_1) \to \pi^b(q_2) + N(p_2)
\end{equation}
can be expressed in terms of four invariant amplitudes $D^\pm$,
$B^\pm$:
\begin{eqnarray}
T_{ab} &=& T^+ \delta_{ab} - T^- i \varepsilon_{abc} \tau_c
\label{piNrel}\\ T^\pm &=& \bar u(p_2) \left[D^\pm(\nu,t) +
\frac{i}{2m_N} \sigma^{\mu\nu} q_{2\mu} q_{1\nu} B^\pm(\nu,t) \right]
u(p_1) \nonumber
\end{eqnarray}
with
\begin{eqnarray}
\begin{array}{llll}
s=& (p_1+q_1)^2 \; \; , & t=& (q_1-q_2)^2 \; \; , \\ u=& (p_1-q_2)^2 \; \;
, & \nu=& \displaystyle \frac{s-u}{4 m_N} \;\; .
\end{array} 
\end{eqnarray}
With the choice of invariant amplitudes $D^\pm$, $B^\pm$, the
low--energy expansion is straightforward: to determine the scattering
amplitude to $O(p^n)$, one has to calculate $D^\pm$ to $O(p^n)$ and
$B^\pm$ to $O(p^{n-2})$.

In the framework of chiral perturbation theory, the first systematic
calculation of pion--nucleon scattering was performed by
\cite{GSS88}. To make the relation between chiral and loop expansions
more transparent than in the original relativistic formulation, the
so--called ``nonrelativistic'' approach of heavy baryon chiral
perturbation theory (HBCHPT) (Jenkins and Manohar 1991; Bernard et
al. 1992) uses a simultaneous expansion in $p/4\pi F_\pi$ and $p/m_N$.

In HBCHPT, the pion--nucleon scattering amplitude is not directly
obtained in the relativistic form (\ref{piNrel}) but rather as (Moj\v
zi\v s 1997)
\begin{equation}
\bar u(p_2) P_v^+ \left[\alpha^\pm + i \varepsilon^{\mu\nu\rho\sigma}
q_{1\mu} q_{2\nu}v_\rho S_\sigma \beta^\pm \right] P_v^+ u(p_1)
\label{piNH}
\end{equation}
$$
v^2=1 \;,\qquad P_v^+=(1+\not\!v)/2 \;, \qquad S^\mu = i/2 \gamma_5
\sigma^{\mu\nu} v_\nu\;.
$$
The HBCHPT amplitudes $\alpha^\pm$, $\beta^\pm$ depend on the choice
of the arbitrary time--like four--vector $v$. A natural and convenient
choice is the initial nucleon rest frame with $v=p_1/m_N$. In this
frame, the relativistic amplitudes are given in terms of the HBCHPT
amplitudes as (Ecker and Moj\v zi\v s 1997)
\begin{eqnarray}
D^\pm &=& \alpha^\pm + \displaystyle \frac{\nu t}{4 m_N} \beta^\pm
\label{INRF}\\ B^\pm &=& - m_N \left(1 - \displaystyle \frac
{t}{4m^2_N} \right) \beta^\pm ~.\nonumber
\end{eqnarray}
Also the amplitudes $D^\pm$, $B^\pm$ will in general depend on the specific
frame associated with $v$. However, since the chiral pion--nucleon 
Lagrangian in HBCHPT can be constructed from a fully
relativistic Lagrangian, the amplitudes  $D^\pm$, $B^\pm$ obtained
from (\ref{INRF}) are guaranteed (Ecker and Moj\v zi\v s 1996) to be 
Lorentz invariant except for terms of at least $O(p^{n+1})$ if the
HBCHPT amplitude
(\ref{piNH}) has been calculated to $O(p^n)$. The same conclusion can
be drawn on the basis of reparametrization invariance (Luke and
Manohar 1992).

The first complete calculation of $\pi N$ scattering to $O(p^3)$ in
the framework of HBCHPT has recently been performed by
\cite{MM97}. The one--loop amplitude of $O(p^3)$ has also been
calculated by \cite{BKM97}, in agreement with \cite{MM97}. Up to
$O(p^2)$, only tree diagrams with vertices from the Lagrangians
of $O(p)$ and $O(p^2)$ appear. In the isospin limit, four
low--energy constants of $O(p^2)$ enter the scattering amplitude in 
addition to the
single coupling constant $g_A$ of the lowest--order Lagrangian. In the
notation of \cite{EM96}, those four dimensionless constants are called
$a_1$, $a_2$, $a_3$ and $a_5$.  The one--loop amplitude
with only lowest--order vertices comes in at $O(p^3)$.
There are at this order also irreducible and reducible tree diagrams 
involving vertices up to $O(p^3)$. In addition to the couplings
already present at $O(p^2)$, five more combinations of low--energy
constants of $O(p^3)$ contribute to the amplitudes denoted 
$b_1+b_2$, $b_3$, $b_6$, $b_{15}-b_{16}$ and $b_{19}$. Thus, the final
renormalized amplitudes depend on $\nu$, $t$, $M_\pi$, $F_\pi$, 
$m_N$, $g_A$ and on 9 combinations of low--energy constants of
$O(p^2)$ and $O(p^3)$. As a technical side--remark, let me
mention that nucleon wave function renormalization in HBCHPT turns out
to be momentum dependent in general (Ecker and Moj\v zi\v s 1997).

The invariant amplitudes $D^\pm$, $B^\pm$ can be projected onto 
partial--wave amplitudes $f_{l\pm}^\pm(s)$ (H\"ohler 1983). 
Threshold parameters are defined as in Eq.~(\ref{effrange}):
\begin{equation}
\Re e\;f_{l\pm}^\pm(s)=q^{2l}\{a_{l\pm}^\pm +q^2 b_{l\pm}^\pm +O(q^4)\}~.
\label{effrpiN}
\end{equation}
To confront the chiral amplitude with experiment, \cite{MM97} has
compared 16 of these threshold parameters with the corresponding
values extrapolated
from experimental data on the basis of the Karlsruhe--Helsinki
phase--shift analysis (Koch and Pietarinen 1980).

Six of the threshold parameters ($D$ and $F$ waves) are independent of
all low--energy constants of $O(p^2)$ and $O(p^3)$. The results are
shown\footnote{I am grateful to Martin Moj\v zi\v s for providing
me with partly unpublished results appearing in Tables
\ref{tab:thresh1}, \ref{tab:thresh2}.}
in Table \ref{tab:thresh1} and compared with \cite{KP80}. 

\begin{table}[t]
\caption{Comparison of two D--wave and four F--wave
threshold parameters up to the first, second and third order
(the two columns differ by higher--order terms) with 
(extrapolated) experimental values (Koch and Pietarinen 1980).
The theoretical values are based on the calculation of \protect\cite{MM97}.
Units are appropriate powers of GeV$^{-1}$.}
\label{tab:thresh1}
\begin{center}
 \begin{tabular}{|c|c|c|c|c|c|} \hline
& & & & & \\
     & \mbox{  } $O(p)$\mbox{  }  &\mbox{  }  $O(p^2)$ \mbox{  } &
   \mbox{  }   $O(p^3)$\mbox{  } & HBCHPT $O(p^3)$ &\mbox{    }
exp.\mbox{    } \\ 
& & & & & \\ \hline
\mbox{  }    $a^+_{2+} $ \mbox{  } &  $0 $ &  $-48 $  & $-35$ &  $ -36 $      
&  $  -36 \pm 7 $ \\ \hline
    $a^-_{2+} $  &  $0 $ &  $48 $  & $56$ &  $ 56 $      
&  $ 64 \pm 3 $ \\ \hline
    $a^+_{3+} $  &  $0 $ &  $0 $  & $226$ &  $ 280 $      
&   \mbox{    }$ 440 \pm 140$ \mbox{    }  \\ \hline
    $a^+_{3-} $  &  $0 $  &  $14 $  & $26$ &  $ 31$      
&  $160 \pm 120$ \\ \hline
    $a^-_{3+} $  &  $0 $      &  $0 $  & $-158$ & $-210 $      
&  $-260 \pm 20\ $ \\ \hline
    $a^-_{3-} $  &  $0 $      &  $-14 $  & $65$ &  $ 57 $   
&  $ 100 \pm 20 $ \\ \hline
  \end{tabular}
\end{center}
\end{table}

Before discussing the results, let me explain the chiral counting in
Tables \ref{tab:thresh1}, \ref{tab:thresh2}. As emphasized before, each
chiral order contains all contributions up to the given order, but it
may also contain some higher--order contributions depending on the
chosen frame (through the choice of the four--vector $v$). The
numerical values in Tables \ref{tab:thresh1}, \ref{tab:thresh2}
 correspond to the initial nucleon rest frame employed
by \cite{MM97}. The column denoted HBCHPT $O(p^3)$ is based on the
amplitudes $\alpha^\pm$, $\beta^\pm$ calculated to $O(p^3)$ in the
initial nucleon rest frame. These amplitudes are then inserted in
Eq.~(\ref{INRF}) to determine the
relativistic amplitudes $D^\pm$, $B^\pm$. The
relations (\ref{INRF}), which are exact to all orders, introduce also
contributions of higher orders than $p^3$. For the column denoted
$O(p^3)$, those higher--order
terms in $D^\pm$, $B^\pm$ have been eliminated. The difference between the
two columns is therefore an indication of the sensitivity of the
respective threshold parameter to higher--order contributions. At
least for the parameters listed in Table \ref{tab:thresh1}, these
differences are within reasonable limits.

The entries of $O(p)$ and $O(p^2)$ in Tables \ref{tab:thresh1} and
\ref{tab:thresh2} differ from those listed in \cite{MM97}. The reason
is that Moj\v zi\v s originally used a different set of invariant
amplitudes ($A^\pm$, $B^\pm$) that are less suited for a proper chiral
counting. 

The main conclusion from Table \ref{tab:thresh1} is a definite
improvement seen at $O(p^3)$. Since there are no low--energy
constants involved (except, of course, $M_\pi$, $F_\pi$, 
$m_N$ and $g_A$), this is clear evidence for the relevance of loop
effects. The numbers shown in Table \ref{tab:thresh1} are based on the
calculation of \cite{MM97}, but essentially the same results have been
obtained by \cite{BKM97}.

\begin{table}[htb]
\caption{Fitted values of ten $\pi N$ threshold parameters up to the first, 
second and third order (the two columns differ by higher--order terms), 
compared with (extrapolated) experimental values (Koch and Pietarinen
1980). The theoretical values are based on the calculation of
\protect\cite{MM97}. Units are appropriate powers of GeV$^{-1}$.}
\label{tab:thresh2}
\begin{center}
  \begin{tabular}{|c|c|c|c|c|c|} \hline
& & & & & \\
     &\mbox{  } $O(p)$ \mbox{  } &\mbox{  }  $O(p^2)$\mbox{  }  &
  \mbox{  }      $O(p^3)$\mbox{  } & HBCHPT $O(p^3)$ &\mbox{    } 
exp.\mbox{    } \\
& & & & & \\ \hline
\mbox{  }    $a^+_0 $ \mbox{  } &  $  0 $      &  $ -0.13 $  & $-0.07$ &  $-0.07 
\pm 0.09$      &  \mbox{  }$ -0.07 \pm 0.01 $ \mbox{  }  \\ \hline
    $b^+_0 $  &  $6.8 $      &  $-15.5 $  & $-11.5$ &  $ -13.9 \pm 3.0$      
&  $-16.9 \pm 2.5 $  \\ \hline
    $a^-_0 $  &  $ 0.55 $      &  $ 0.55 $  & $0.67$ & 
$ 0.67 \pm 0.10 $      &  $ 0.66 \pm 0.01$  \\ \hline
    $b^-_0 $  &  $2.6 $      &  $2.6 $  & $6.7$ &  $ 5.5 
\pm 6.7 $      &  $ 5.1 \pm 2.3 $   \\ \hline
    $a^+_{1+} $  &  $15.4 $      &  $48.4 $  & $49.6$ &  $ 50.4 
\pm 1.1 $      &  $ 50.5 \pm 0.5 $  \\ \hline
    $a^+_{1-} $  &  $-37.5 $      &  $-4.5 $  & $-22.4$ &  $-21.6 \pm 1.8 $ 
     &  $-21.6 \pm 0.5 $    \\ \hline
    $a^-_{1+} $  &  $-15.4 $      &  $-27.5  $  & $-31.4$ &  $-31.0 
\pm 0.8 $      &  $-31.0 \pm 0.6 $    \\ \hline
    $a^-_{1-} $  &  $-15.5 $      &  $8.6 $  & $-4.9$ &  $ -4.5 
\pm 1.0 $      &  $ -4.4 \pm 0.4 $  \\ \hline
    $a^+_{2-} $  &  $-4.4 $      &  $6.9 $  & $28.8$ &  $ 31.2 
\pm 0.3 $      &  $ 44 \pm 7  $  \\ \hline
    $a^-_{2-} $  &  $4.4 $      &  $-12.8 $  & $-0.1$ &  $ -5.0 
\pm 0.2  $      &  $ 2 \pm 3 $  \\ \hline
   \end{tabular}

\end{center}

\end{table}

The remaining 10 threshold parameters do depend on the 9 low--energy
constants of $O(p^2)$ and $O(p^3)$. To fit these 9 constants,
\cite{MM97} has included the nucleon $\sigma$--term and the  $\pi N$
coupling constant $g_{\pi N}$ that depend on the same constants. The
results of the fit for the threshold parameters are shown in Table
\ref{tab:thresh2}. Two immediate conclusions are:
\begin{itemize}
\item
Not too surprisingly with 9 parameters for 12 observables, the
``experimental'' values can be reproduced with the chiral amplitude to
$O(p^3)$. Incidentally, the fitted value of the $\sigma$--term tends
to be larger (Moj\v zi\v s 1997) than the canonical value (Gasser et
al. 1991). The discrepancy becomes smaller when increasing the errors
for the threshold parameters.
\item
In many cases, the corrections of $O(p^3)$ are sizable and definitely
bigger than what naive chiral order--of--magnitude estimates would
suggest. For instance, at threshold the third--order corrections
$\alpha_3^-$, $\beta_3^+$ due to the low--energy constants $b_i$ of
$O(p^3)$ are some 30 \% of the leading--order amplitudes.
\end{itemize}

At least as interesting as the fitted values of the threshold
parameters are the corresponding values of the low--energy constants 
shown in Table \ref{tab:LECS}. The first observation is that most of
these constants are of $O(1)$ as expected from naive chiral
dimensional analysis. Looking a little closer into the calculation of
\cite{MM97}, one finds that in some amplitudes the low--energy
constants of $O(p^3)$ interfere constructively near threshold so that
their overall effect is larger than the naive estimate.

The values for the low--energy constants $a_i$ in Table \ref{tab:LECS}
agree very well with an independent analysis (although some of the
input data are the same) of \cite{BKM97}. Moreover, these authors have
shown that the specific values of $a_1$, $a_2$ and $a_5$
can be understood as being mainly due
to $\Delta$ and $\rho$ exchange, whereas $a_3$ (appearing in the $\sigma$
term) is compatible with scalar resonance dominance. I conclude
that the low--energy constants of $O(p^2)$ in the pion--nucleon
Lagrangian are under good control, both numerically and conceptually.
A similar analysis is not yet available for the constants $b_i$.

\begin{table}[htb]
\caption{Values of low--energy constants of $O(p^2)$ and $O(p^3)$ 
from fitting (Moj\v zi\v s 1997) ten $\pi N$
threshold parameters,  the nucleon $\sigma$--term and 
the Goldberger--Treiman discrepancy. The $\widetilde{b}_i$ are
scale independent versions of the $b_i$.}
\label{tab:LECS}

\begin{center}
  \begin{tabular}{|c|c|} \hline
     $a_1$  &  $-2.60 \pm 0.03 $ \\ \hline
     $a_2$  &  $\quad 1.40\ \pm 0.05$  \\ \hline
     $a_3$  & \mbox{  } $-1.00\ \pm 0.06$ \mbox{  } \\ \hline
     $a_5$  &  $\quad 3.30 \pm 0.05 $ \\ \hline \hline
     $\widetilde{b}_1+\widetilde{b}_2$      &  $\quad 2.4\ \pm 0.3 $ \\ \hline
     $\widetilde{b}_3$       &  $-2.8\ \pm 0.6 $  \\ \hline
     $\widetilde{b}_6$       & $\quad 1.4\ \pm 0.3 $  \\ \hline
   \mbox{  }  $b_{16}-\widetilde{b}_{15}$  \mbox{  }  
     & $\quad 6.1\ \pm 0.6 $  \\ \hline
     $b_{19}$       & $ -2.4  \ \pm 0.4  $ \\ \hline
  \end{tabular}
\end{center}

\end{table}

\section{Summary and Outlook}

\subsection{$\pi\pi\to\pi\pi$}
Unlike for
most other processes discussed during this Workshop, the available CHPT
calculations to $O(p^6)$ are amply sufficient, even in view of the
forthcoming precision experiments. However, while waiting for the
results from KLOE at DA$\Phi$NE (Lee-Franzini 1997), DIRAC at CERN
(Schacher 1997) and E865 at BNL (Lowe 1997), several
things remain to be done. Among the most interesting topics are the
following: 
\begin{itemize}
\item
There is a complementarity between the field theoretic calculation
(Bijnens et al. 1996, 1997) and the dispersive one (Knecht et al.
1995, 1996). While the first method fully accounts for the infrared
structure, the latter encompasses the high--energy
information via sum rules for the relevant parameters. It
remains to bring the two ingredients together in an optimal way
through a Roy--type analysis (Colangelo et al. 1997b) to extract
especially the $S$--wave scattering lengths $a_0^0$, $a_0^2$ and the
low--energy constants $\bar{l}{_1}$, $\bar{l}{_2}$ with reliable
errors. 
\item
Isospin violation and electromagnetic corrections have to be
included. First results are already available (Mei\ss ner et
al. 1997; Knecht and Urech 1997).
\end{itemize}

Concerning the $I=0$ $S$--wave scattering length, I conclude
that a value
\begin{equation}
a_0^0 = 0.21 \div 0.22
\end{equation}
is well established on the basis of existing CHPT calculations in the
standard scheme.
Such a value is in perfect agreement with the available experimental
information. Thus, there is at present no indication from pion--pion
scattering against the standard scenario of chiral symmetry breaking
with a dominant quark condensate.

\subsection{$\pi N \to\pi N$}
The first complete analysis of $\pi N$ scattering to $O(p^3)$ by 
\cite{MM97} has produced very encouraging
results. However, we are still far from the theoretical precision
attained in $\pi\pi$ scattering. Among the most urgent tasks are the
following: 
\begin{itemize}
\item
The chiral amplitude should be confronted with extrapolated and/or
real phase shifts to check the range of validity of the chiral
expansion and to control the stability of the low--energy constants
involved. In a recent paper that has appeared after the Workshop,
\cite{ET97} have actually calculated the phase shifts to $O(p^3)$ and
compared them with experiment, using a somewhat different approach 
that is claimed to be equivalent to HBCHPT.
\item
The threshold parameters of \cite{KP80} are derived from a data sample
parts of which are claimed to be inconsistent (Pavan 1997). An update on
the basis of generally accepted experimental input would be highly
welcome for meaningful tests of chiral perturbation theory.
\item
As the low--energy constants of $O(p^2)$ are now well understood, both
phenomenologically and theoretically, a similar analysis for the
constants of $O(p^3)$ is called for. The results should eventually be
contrasted with an alternative analysis based on the
$\varepsilon$--expansion with
the $\Delta$--isobar as explicit degree of freedom (Kambor 1997).
\item
Finally, in spite of the encouraging results to $O(p^3)$, the $O(p^4)$
calculation is absolutely necessary to complete the pion--nucleon
amplitude to one--loop accuracy.
\end{itemize}

\section*{Acknowledgements}
I am indebted to many friends and colleagues for sharing with me their
knowledge of the matters reported here, especially to Hans Bijnens,
Gilberto Colangelo, J\"urg Gasser, Marc Knecht, Heiri Leutwyler,
Martin Moj\v zi\v s, Mikko Sainio and Jan Stern. For the efficient
organization of the Workshop, I want to thank Aron Bernstein, Dieter
Drechsel and Thomas Walcher.

\end{document}